\documentclass{PoS}
\usepackage{wrapfig}
\usepackage{rotating}
\usepackage{multirow}

\title{{\small{DESY 16-185, DO-TH 16/26}}
\\
PDFs, $\alpha_s$, and quark masses from global fits}

\ShortTitle{PDFs, $\alpha_s$, and quark masses from global fits}

\author{\speaker{Sergey Alekhin}%
         \thanks{This work was supported in part by the
European Commission through contract PITN-GA-2012-316704 ({HIGGS\-TOOLS}).
}\\ 
II. Institut f\"ur Theoretische Physik, Universit\"at Hamburg,
    Luruper Chaussee 149, D-22761 Hamburg, Germany;\\
        Institute for High Energy Physics,142281 Protvino, Russia\\
        E-mail: \email{sergey.alekhin@desy.de}}

\author{Johannes Bl\"umlein\\ 
        Deutsches Elektronensynchrotron DESY, Platanenallee 6, D--15738 Zeuthen, Germany\\
        E-mail: \email{Johannes.Bluemlein@desy.de}}

\author{Sven-Olaf Moch\\
II. Institut f\"ur Theoretische Physik, Universit\"at Hamburg,
    Luruper Chaussee 149, D-22761 Hamburg, Germany \\      
 E-mail: \email{sven-olaf.moch@desy.de}}

\author{Ringaile Pla\v cakyt\. e \\
   Deutsches Elektronensynchrotron DESY, 
   Notkestra{\ss}e 85, D--22607 Hamburg, Germany\\
        E-mail: \email{ringaile.placakyte@desy.de}}

\abstract{The strong coupling constant $\alpha_s$ 
and the heavy-quark masses, $m_c$, $m_b$, $m_t$ are extracted 
simultaneosly with the parton distribution functions (PDFs) in the 
updated ABM12 fit including recent data from CERN-SPS,
HERA, Tevatron, and the LHC. The values of 
\begin{eqnarray}
\nonumber \alpha_s(M_Z)&=&0.1147\pm0.0008~({\rm exp.)},\\  \nonumber 
m_c(m_c)&=&1.252\pm 0.018~({\rm exp.})~{\rm GeV},\\ \nonumber
m_b(m_b)&=&3.83\pm0.12~({\rm exp.})~{\rm GeV},\\ \nonumber
m_t(m_t)&=&160.9\pm1.1~({\rm exp.})~{\rm GeV}
\end{eqnarray}
are obtained with the $\overline{MS}$ heavy-quark mass definition 
being employed 
throughout the analysis. 
}

\FullConference{Loops and Legs in Quantum Field Theory\\
		24-29 April, 2016\\
		Leipzig, Germany}

\begin{document}
\begin{figure}
\begin{minipage}[c]{0.62\textwidth}
\includegraphics[width=0.85\textwidth, angle=0]{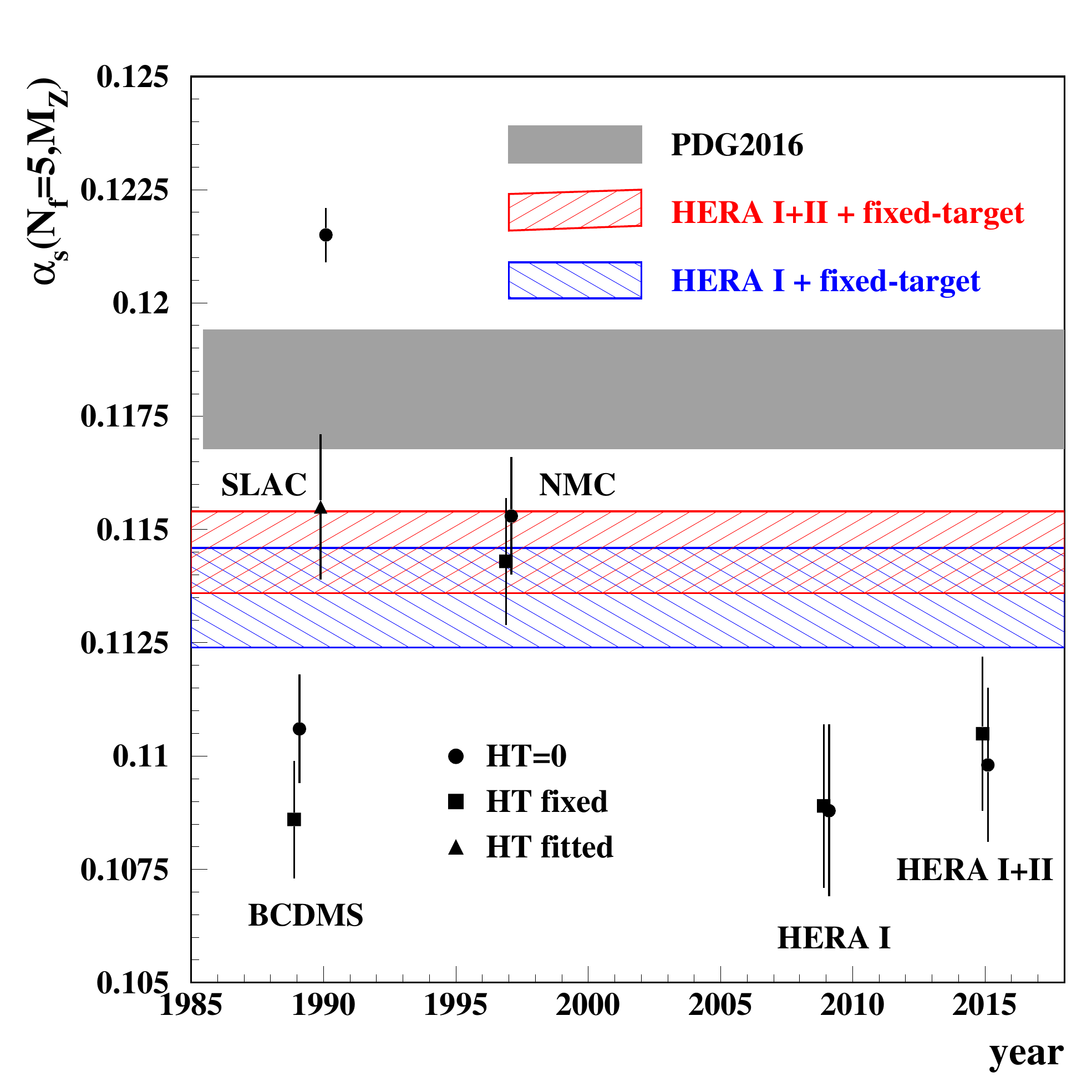}
  \end{minipage}\hfill
  \begin{minipage}[c]{0.36\textwidth}
\caption{\small
The value of $\alpha_s(M_Z)$ preferred by various DIS data samples 
employed in the 
present analysis w.r.t. the year of the data publication. Three variants
of the fit with different treatment of the HT terms are presented: The
HT set to 0 or to the ones obtained in the combined fit 
(circles and squares, respectively) and fitting to the one particular data set 
(triangles). 
The $\alpha_s$ bands obtained by using the combination of the fixed-target SLAC, BCDMS,
and NMC samples with those from HERA Run~I (left-tilted hatches) and 
Run~I+II (right-tilted hatches), as well as the PDG 
average~\cite{Agashe:2014kda}, are given for 
comparison.  }
\label{fig:alps}
\end{minipage}
\end{figure}

With the form of the Standard-Model (SM) Lagrangian widely validated after
the experimental discovery of the Higgs boson its 
fundamental parameters play increasing role in search of physics beyond 
SM and related topics in cosmology~\cite{Shaposhnikov:2009pv}. These parameters 
can be extracted from the various experimental data 
with an accuracy of O(1\%) achieved for the 
heavy-quark masses and the strong coupling 
constant $\alpha_s$~\cite{Agashe:2014kda}. 
The latter was obtained in particular in the ABM12 fit~\cite{Alekhin:2013nda}
aimed to determine the parton distribution functions (PDFs) from the 
combination of
lepton-nucleon deep-inelastic-scattering (DIS) and Drell-Yan (DY)
data. The value of $\alpha_s$ obtained in this way is mainly driven by 
the DIS sample, while the DY part is employed to facilitate 
the disentangling of the quark species. 
The input for the ABM12 fit has been recently updated 
with the latest DIS and DY data from CERN-SPS, HERA, Tevatron, and the 
LHC~\cite{Alekhin:2016uxn,Alekhin:2015cza,Alekhin:2014sya}.
The main impact of this input on $\alpha_s$ stems from the new inclusive 
DIS data set representing a Run~I+II combination of the results obtained by the 
H1 and ZEUS experiments during the whole period of the HERA collider operation. 
This tendency
was checked by performing a variant of the present analysis with no other DIS 
data included. The value of $\alpha_s$ obtained in this way is displayed in 
Fig.~\ref{fig:alps} in comparison with the one preferred by the earlier 
Run~I combination of the H1 and ZEUS data accumulated in the first HERA run. 
The Run~I+II value of $\alpha_s$ moves up by $1\sigma$ which causes a 
corresponding shift in the results of the combined analysis including 
all DIS data, although the combined value of 
\begin{equation}
\alpha_s(M_Z)=0.1145\pm0.0009~({\rm exp.})
\label{eq:alps}
\end{equation}
obtained is still
lower than the PDG average, cf. Fig.~\ref{fig:alps}. Other DIS experiments 
included into our analysis, SLAC, BCDMS and NMC, also prefer relatively small
$\alpha_s$ values, however, for the case of SLAC it is achieved only if the 
higher-twist (HT) terms are taken into account properly. 
The BCDMS, NMC, and HERA Run~I+II data are less sensitive to HT 
terms, while the HERA Run~I ones are not sensitive at all.  
Indeed, in the region of small $x$, which is 
controlled by the HERA Run~I data in the ABM12 fit, the HT values obtained 
are fully negligible 
within the uncertainties and in the present analysis a
non-zero HT contribution to the longitudinal structure 
function $F_L$ is found due to the better accuracy of the Run~I+II 
combination~\cite{Alekhin:2016uxn,Harland-Lang:2016yfn,Abt:2016vjh}.

\begin{sidewaysfigure}[th!]
\centerline{
  \includegraphics[width=10.0cm]{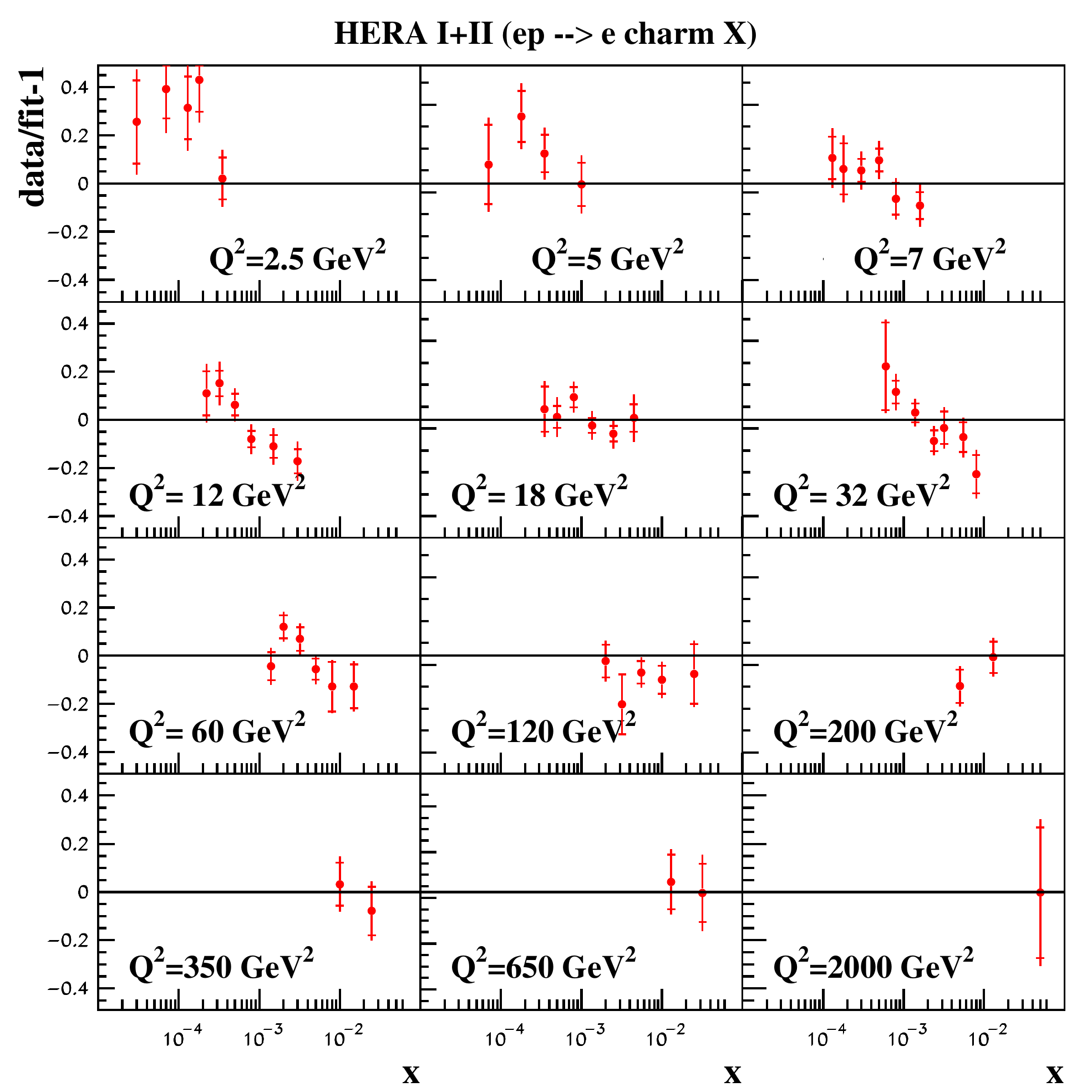}
  \includegraphics[width=10.0cm]{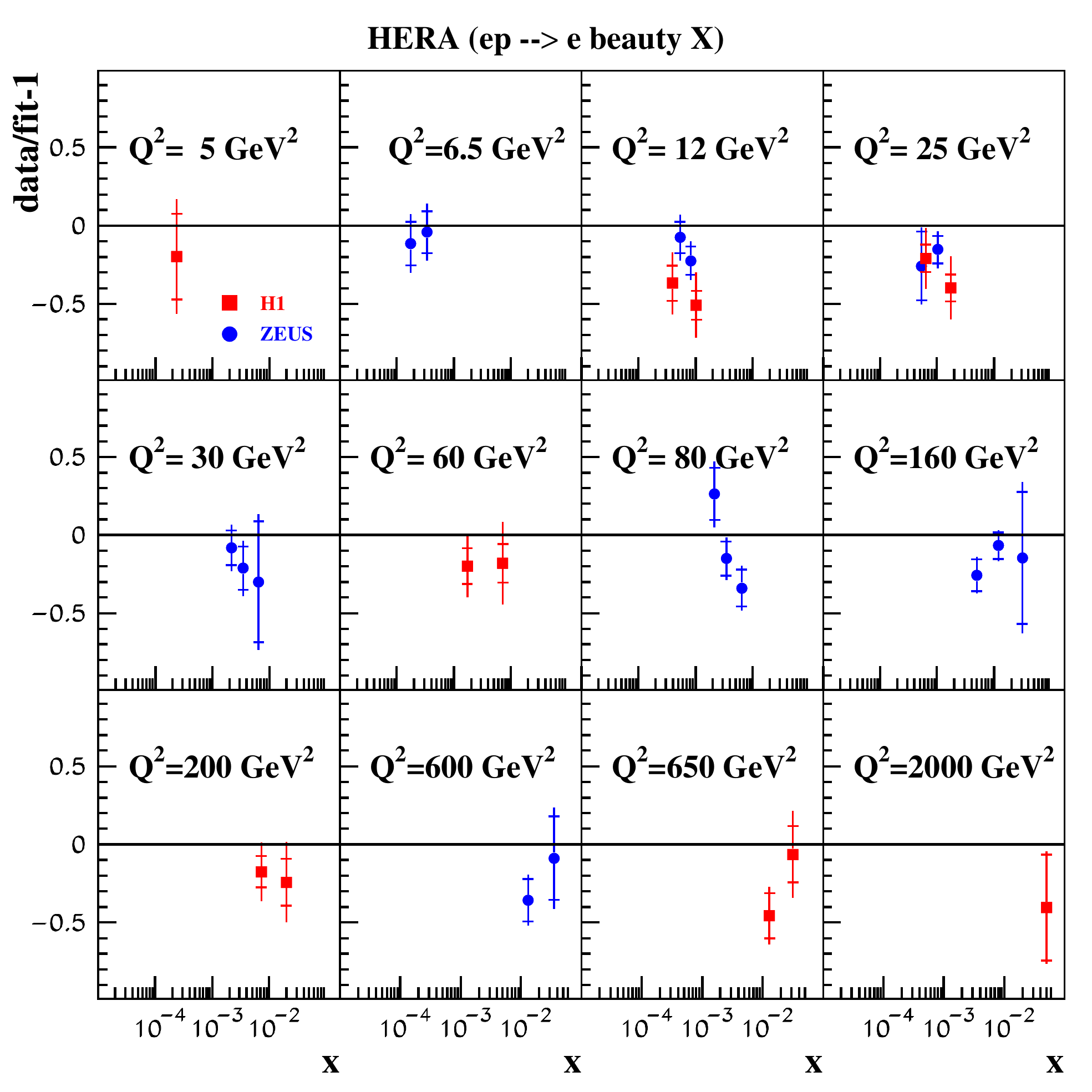}}
  \caption{\small
    \label{fig:hera-ccbar}
Left panel:  The pulls versus Bjorken $x$ 
in bins of the momentum transfer $Q^2$ for the 
combination of NC DIS inclusive charm-quark production data
from the H1 and ZEUS experiments at 
HERA~\cite{Abramowicz:1900rp}.
Right panel: The same for the separate bottom-quark
    production data from H1~\cite{Aaron:2009af} (squares)
and ZEUS~\cite{Abramowicz:2014zub} (circles).
  }
\end{sidewaysfigure}

In addition to the inclusive HERA sample we also employ the data on 
the $c$- and $b$-quark DIS production, cf. Fig.~\ref{fig:hera-ccbar}.
They span the region of momentum transfer $Q^2$ down to few ${\rm GeV}^2$ and 
therefore are sensitive to the values of heavy quark masses $m_{c,b}$.  
In our analysis the DIS process is described within the 
3-flavour fixed-flavor-number (FFN)
factorization scheme with the QCD corrections up to NNLO
both for the massless and massive contributions taken into account. 
The NNLO terms in the neutral-current massive Wilson coefficients
are derived from the high-energy asymptotic expressions and the 
threshold-resummation results matched using available
NNLO terms in the massive operator matrix elements 
(OMEs)~\cite{Ablinger:2014nga,Behring:2014eya}.
For the first time such a parameterization~\cite{Kawamura:2012cr}
was used in the ABM12 analysis. In the present analysis we take an improved
form of the parameterization~\cite{Kawamura:2012cr} employing 
the NNLO pure-singlet terms in the
massive OMEs~\cite{Ablinger:2014nga}.
The NNLO corrections to charged-current $c$-quark production 
are taken in the asymptotic 
form~\cite{Blumlein:2014fqa,Alekhin:2014sya} valid for $Q^2\gg m_c^2$ 
which are therefore applied only to the HERA data occupying the region of
$Q^2>300~{\rm GeV}^2$.
All massive terms in the scheme adopted are considered in
the $\overline{MS}$-scheme for the heavy-quark mass since this approach 
was shown to provide better perturbative stability if compared to the 
pole-mass case~\cite{Alekhin:2010sv}. 
Furthermore, the fit based on this  
framework provides a good agreement with the HERA data 
on the heavy-quark production up to the largest values of $Q^2$
available, cf. Fig.~\ref{fig:hera-ccbar}.
The $\overline{MS}$-values of the $c$- and $b$-quark masses 
are determined from the fit simultaneously with the PDFs as
\begin{equation}
m_c(m_c)=1.252\pm 0.018~({\rm exp.})~{\rm GeV},~~~~~~~~~~
m_b(m_b)=3.83\pm0.12~({\rm exp.})~{\rm GeV}.
\label{eq:masses}
\end{equation}
They are in a good agreement with other NNLO determinations
based on a variety of different experimental data~\cite{Agashe:2014kda}.
On the other hand, a value of the $c$-quark pole mass 
$m_c^{pole}\sim~1.3~{\rm GeV}$ is 
commonly set in the PDF fits based on the variable-flavor-number (VFN)
factorization scheme~\cite{Dulat:2015mca,Harland-Lang:2014zoa,Ball:2014uwa}. 
This setting is dramatically different from ours since the value of 
$m_c(m_c)$ Eq.~(\ref{eq:masses}) corresponds to 
$m_c^{pole}=2.4~{\rm GeV}$ with account of the 4-loop corrections 
to the matching relations~\cite{Marquard:2015qpa}.
We note that the values of $\chi^2$ obtained for the HERA inclusive and 
semi-inclusive data in the VFN-based PDF fits are in general larger
than the FFN ones. This also demonstrates the benefit of 
the FFN approach~\cite{Accardi:2016ndt}.

\begin{figure}[th!]
\centerline{
  \includegraphics[width=16.0cm]{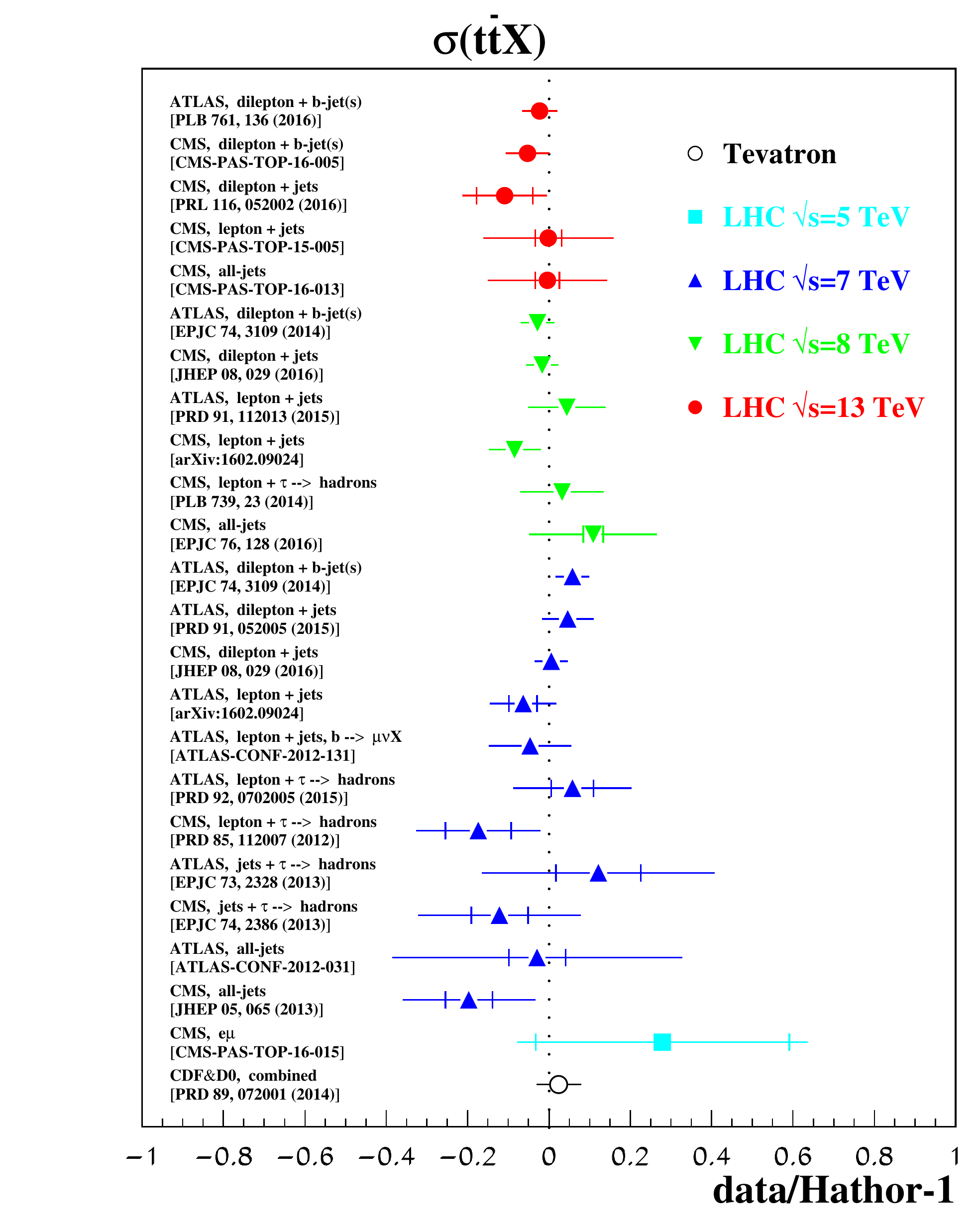}}
  \caption{\small
    \label{fig:ttbar-pulls}
    The pulls for data on top-quark pair production from the Tevatron (CDF and D0)
    at $\sqrt s = 1.96$~TeV 
    and the LHC (ATLAS and CMS) at $\sqrt s = 5, 7, 8$ and $13$~TeV 
    with respect to our NNLO fit.
    The NNLO QCD predictions have been obtained with 
Hathor~\cite{Aliev:2010zk}. 
  }
\end{figure}

\begin{figure}[th!]
\centerline{
  \includegraphics[width=16.0cm]{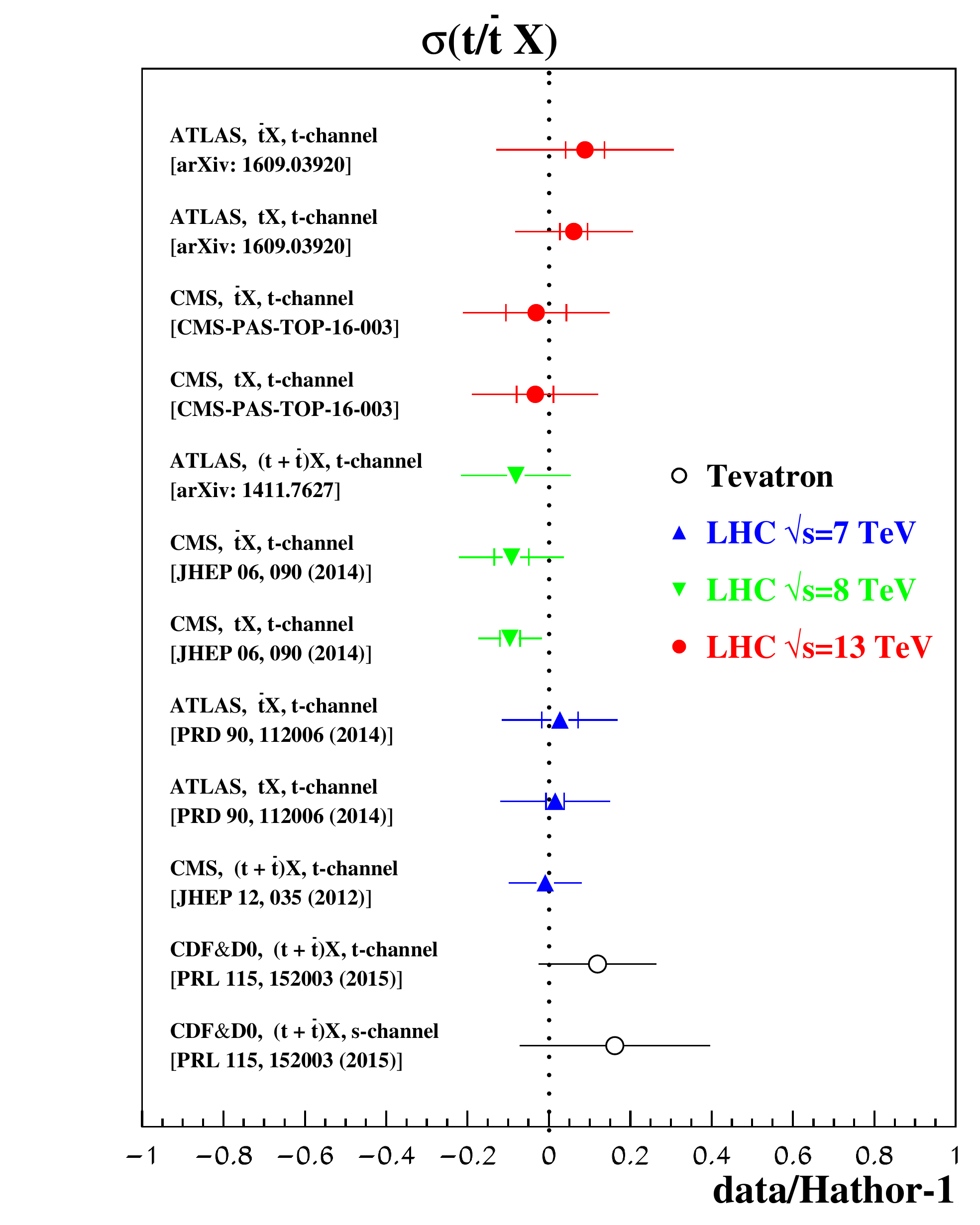}}
  \caption{\small
    \label{fig:single-top-pulls}
    Same as Fig.~{\protect \ref{fig:ttbar-pulls}} 
for the data on single-top production 
    in the $s$- and $t$-channel.
    The NNLO QCD predictions have been computed with Hathor~\cite{Kant:2014oha} 
    as described in text.
  }
\end{figure}

The data on single-top and $t\bar{t}$ hadro-production cross sections 
from Tevatron and the LHC included in the present
analysis allow to determine the $t$-quark 
mass $m_t$ once it is considered as a fit parameter. The value of $m_t$
 determined in such a way is a well-defined quantity, in contrast to 
experimental determinations commonly sensitive to details of Monte-Carlo 
modeling. Similarly to the case of $c$- and $b$-quark DIS production we employ 
the $\overline{MS}$ definition for the $t$-quark hadro-production that 
provides better perturbative stability here again~\cite{Alekhin:2013nda}.
For this purpose we run the Hathor code~\cite{Aliev:2010zk,Kant:2014oha}
with the full NNLO corrections to the $t\bar{t}$~\cite{Czakon:2013goa} and
the $t$-channel single-top production~\cite{Brucherseifer:2014ama}, 
while the NNLO terms for the $s$-channel single-top production 
are computed using the threshold-resummation 
approximation~\cite{Alekhin:2016jjz}. The $t$-quark data collected in a wide 
range of the c.m.s. energy are well accommodated in the fit, cf.
Figs.~\ref{fig:ttbar-pulls},{\protect \ref{fig:single-top-pulls}}. The  
value of 
\begin{equation}
m_t(m_t)=160.9\pm1.1~({\rm exp.})~{\rm GeV}
\end{equation}
preferred by the data 
is in a broad agreement with other determinations based on the 
$t$-production cross section~\cite{Agashe:2014kda} and the value of 
$m_t(m_t)=158.9\pm3.4~({\rm exp.})~{\rm GeV}$ 
extracted from the single-top production data only and the ABM12 PDFs 
used~\cite{Alekhin:2016jjz}. The value of 
$\alpha_s(M_Z)=0.1147\pm0.0008~({\rm exp.)}$ 
obtained in the latter variant of the fit 
with the $t$-quark data included comes out somewhat bigger 
than the value of Eq.~(\ref{eq:alps}), 
however, the difference lays within the uncertainties.

\end{document}